\documentstyle[12pt,epsf]{article}
\begin{document}
\hoffset=-1.2cm
\hsize=16cm
\vsize=24cm

{\centerline{\bf {A STUDY OF SCHWINGER-DYSON EQUATIONS FOR}}}
\vskip 2mm
{\centerline{\bf {YUKAWA AND WESS-ZUMINO MODELS}}}
\vskip 1cm
\baselineskip=7mm
{\centerline{{\bf{A. Bashir}}}}
\vskip 5mm
{\centerline{Instituto de F{\'\i}sica y Matem\'aticas}}
{\centerline{Universidad Michoacana de San Nicol\'as de Hidalgo}}
{\centerline{Apdo. Postal 2-82 , Morelia}}
{\centerline{Michoac\'an, M\'exico}}
\vskip 1cm
{\centerline{{\bf{J. Lorenzo Diaz-Cruz}}}}
\vskip 5mm
{\centerline{Instituto de F{\'\i}sica ``Luis Rivera Terrazas"}}
{\centerline{Benem\'erita Universidad Aut\'onoma de Puebla}}
{\centerline{Ap. Postal J-48, 72500 Puebla}}
{\centerline{Puebla, M\'exico}}
\vskip 2cm
{\centerline {ABSTRACT}}
{\noindent We study Schwinger-Dyson equation
for fermions in Yukawa and Wess-Zumino models, in terms of dynamical 
mass generation and the wavefunction renormalization function.
In the Yukawa model with 
$\gamma_5$-type interaction between scalars and fermions,
we find a critical coupling in the quenched approximation
above which fermions acquire dynamical mass. This is shown to be
true beyond the bare 3-point vertex approximation. 
In the Wess-Zumino 
model \cite{wess}, there is a neat 
cancellation of terms leading to no dynamical mass for fermions.
We comment on the conditions under which these results are
general beyond the
rainbow approximation and also on the ones under which supersymmetry is
preserved and the scalars as well do not acquire mass. 
The results are in accordance with the non-renormalization theorem
at least to order $\alpha$ in perturbation theory. In both
the models, we also evaluate the
wavefunction renormalization function, analytically in the neighbourhood 
of the critical coupling and numerically, away from it. }

\vfil\eject
\vskip 2cm
\section{Introduction}
\baselineskip=7.5mm

Despite the success of quantum field theory in the description of
the behaviour of elementary particles in the perturbative regime
of interactions, it still remains a challenge to understand
the non-perturbative domain satisfactorily. One of the methods which has
gained attention in this regard in recent years is the study of
Schwinger-Dyson equations (SDEs) \cite{roberts}. Despite the difficulties involved in 
finding a non-perturbative truncation of these equations, this approach 
has been very successful in addressing issues like dynamical
mass generation for fundamental fermions when they are involved in
sufficiently strong interactions \cite{miransky}. Moreover, recent 
attempts, e.g., \cite{mike,dong,bashir} to
improve the reliability of the approximations used have increased the
credibility of the results obtained through such studies.

Application of Schwinger-Dyson formalism to supersymmetric (SUSY) models has
been less extensive. In supersymmetric Quantum Electrodynamics
(SQED), based upon the arguments of non-renormalization theorem and
gauge invariance, it is expected to be impossible to obtain dynamical
mass generation for fermions \cite{clark} though some studies 
\cite{kaiser} argue
that it is probably possible to break chiral symmetry dynamically in
SQED. We postpone the study of SQED for a future work. In this paper,
we take the simplest SUSY model, i.e the Wess-Zumino model
and attempt to solve the corresponding Schwinger-Dyson equations for the
fermion and scalar propagators.
We believe this exercise will provide us with a deeper insight
into how the role of supersymmetry in the context of dynamical
mass generation translates into the language of Schwinger-Dyson
equations. 

Such a study should provide us with a better starting 
point for more complicated  SUSY theories such as SQED and SQCD.
In the latter theory, a need also exists to further explore connections 
between the Holomorphic approach and that of the Schwinger-Dyson 
equations \cite{apple}.

We first study the Yukawa model with one real scalar and one Majorana
fermion, which can be considered as a truncated Wess-Zumino model.
We discuss this model in some detail for two reasons, first being that
it is interesting in its own right because, after all, it is Yukawa
interactions which are responsible for giving masses to fermions
in the Standard Model (SM). Secondly, extending the Yukawa model
by doubling the scalar degrees of freedom provides us with a clear 
understanding of how supersymmetry works. We use the quenched approximation.
Keeping in mind the perturbative expansion of the 3-point vertex
beyond the lowest order and its transformation under charge conjugation
symmetry, we propose an {\it ansatz} for the full effective vertex.
One of the advantages of using this vertex is that the equations for the mass 
function ${\cal M}(p^2)$ and the wavefunction renormalization $F(p^2)$
decouple completely in the neighbourhood of the critical 
coupling, $\alpha_c$ above which mass is generated for the fermions, 
and partly above it. 
We solve both the equations to find 
analytical expressions for $F(p^2)$ and the anomalous mass dimensions
in the neighbourhood of $\alpha_c$. The results show that non-perturbative 
interaction
of fermions with fundamental scalars can give masses to fermions in
a dynamical way provided the interaction is strong enough.
We use numerical calculation to draw Euclidean mass of the fermions
as a function of the coupling, and confirm that it obeys Miransky scaling.
We also evaluate $F(p^2)$ numerically.
We then extend the particle spectrum by doubling the number of scalars
and imposing relations for the couplings that define the Wess-Zumino
model. Due to the presence of the additional symmetry, we are able to
extract useful information beyond the rainbow approximation.


\section{The Yukawa Model}

Consider a massless Lagrangian with one Majorana fermion and one real
scalar interacting with each other through a $\gamma_5$-type interaction:
\vspace {2mm}
\begin{eqnarray}
{\cal{L}} = \frac{1}{2} (\partial_{\mu} A)^2 + 
   \frac{1}{2} i (\bar{\psi} \gamma^{\mu} \partial_{\mu} \psi)
 -\frac{1}{2} g^2 A^4 + ig\bar{\psi} \gamma_5 \psi	A \;.
\end{eqnarray}
The corresponding Schwinger-Dyson equation for the fermion propagator, 
$S_{F}(p)$, is displayed in Fig. 1. Motivated by the success of 
the quenched approximation in QED and QCD, we neglect the fermion loops. 
Moreover, as a first step towards truncating the infinite set of 
Schwinger-Dyson equations, we drop all 4-point functions. The full scalar
propagator can then be replaced by its bare counterpart. Using Feynman
rules, the Schwinger-Dyson equation can be written as:
\vspace {2mm}
\begin{eqnarray}
     -iS_{F}^{-1}(p)\,=\,-iS_{F}^{0^{-1}}(p)\,-\,\int \frac{d^4k}{(2\pi)^4}\,
  (-2g  \gamma_{5}) \; (iS_{F}(k)) \; (-2g\gamma_5 \Gamma_A(k,p)) \; 
(\frac{i}{q^2})    \quad ,
\end{eqnarray}
where  $q=k-p$ and $S_{F}(p)$ can be expressed in terms of two Lorentz scalar
functions, $F(p^2)$, the wavefunction renormalization and
${\cal M}(p^2)$, the mass function, so that
\vspace {2mm}
\begin{eqnarray}
    S_{F}(p)&=& \frac{F(p^2)}{\not \! p- {\cal M}(p^2)}\qquad .
\end{eqnarray}   
The bare propagator $ S_{F}^{0}(k)= 1/ (\rlap / p ) $, where 
the bare mass has been taken to be zero. 
We can project out equations for 
$F(p^2)$ and ${\cal M}(p^2)$
by taking the
trace of Eq. (2) having multiplied by $\not \! p$ and 1 in turn.
On Wick rotating to Euclidean space,
\vspace {2mm}
\begin{eqnarray}
 F(p^2) &=& 1 \;- \frac{\alpha}{\pi^3} \frac{1}{p^2} \int
                          d^4k \,
                          \frac{F(k^2) F(p^2)}{k^2+{\cal M}^2(k^2)}\
                          \frac{k \cdot p}{q^2}\ \; \Gamma_A(k,p)   
			  \label{Feqn}     \\    
 {\cal M}(p^2)  &=& \frac{\alpha}{\pi^3} \int 
                          d^4k \,
                          \frac{{\cal M}(k^2)}{k^2+{\cal M}^2(k^2)}\ 
                          \frac{F(k^2)F(p^2)}{q^2} \; \Gamma_A(k,p) \label{Meqn}
\end{eqnarray}
where $\alpha=g^2/ 4\pi$.

It is here that we cannot proceed any further without making an 
{\it ansatz} for $\Gamma_A(k,p)$. Any {\it ansatz} for the 3-point vertex
must fulfill at least the following requirements:
\begin{itemize}
\item   Perturbatively, we must have
        $\Gamma_A(k,p)=1 + {\cal O}(g^2)$.
\item   It must be symmetric in $k$ and $p$.
\end{itemize}
Moreover, as the SDEs relate the 2-point function with the 3-point
function, the expression for the full vertex is expected to 
involve functions $F(p^2)$ or/and ${\cal M}(p^2)$.

        The commonly used ansatz in the non-perturbative study of the
SDEs is the bare vertex ansatz. For the Yukawa Model under discussion,
it implies $\Gamma_A(k,p)=1$. It agrees with the lowest order 
perturbation theory. The only truncation of the complete set of 
Schwinger-Dyson equations known so far that avoids any assumptions 
other than the smallness of the coupling at every level of this
approximation
is the perturbation theory. Therefore, it is natural to
assume that physically meaningful solutions
of the Schwinger-Dyson equations must agree with perturbative results
in the weak coupling regime. It requires, e.g., that every non-perturbative
{\em ansatz} chosen for the full vertex must reduce to its perturbative
counterpart when the interactions are weak. Bare vertex fulfills this
requirement to the lowest order in perturbation theory. Any other vertex 
which fulfills this condition and does not violate other requirements
is at least as good as the bare vertex. One of the simplest non-perturbative
vertices can be constructed by realizing that Eq.(\ref{Feqn}) yields 
the following expansion of $F(p^2)$ in perturbation theory:
\begin{eqnarray}
F(p^2)=1+ {\cal O}(\alpha) 
\end{eqnarray}
Therefore, a simple candidate for the 3-point
vertex can be written as:
\vspace {2mm}
\begin{eqnarray}
   \Gamma_A(k,p)= \frac{1}{F(k^2)F(p^2)} \qquad.
\end{eqnarray}
Perturbatively, it gives  
\begin{eqnarray}
\Gamma_A(k,p)= \frac{1}{ [1 + {\cal O}(\alpha)] [1 + {\cal O}(\alpha)] }
= 1 + {\cal O}(\alpha)
\end{eqnarray}
Therefore, to the lowest order in perturbation theory, our vertex
{\em ansatz} reduces to the bare vertex. 

      It is exceedingly complicated to ensure that at higher orders,
the non-perturbative vertex reduces to its perturbative counterpart 
in the weak coupling regime. We do not aim at it in this paper. However,
we demonstrate that even up to next to lowest order, i.e, to 
${\cal O}(\alpha)$, our {\em ansatz} is correct to the extent that
both the {\em ansatz} and the real vertex have the logarithmically
divergent behaviour in the ultraviolet regime.

       Fig. 2 represents the perturbative expansion of the 
3-point fermion-scalar vertex to ${\cal O}(\alpha)$. Using
Feynman rules, we can write it as follows:
\begin{eqnarray}
 -2g \gamma_5 \Gamma_A = -2g \gamma_5 + \int \frac{d^4 w}{(2 \pi)^4}
 \, (-2g \gamma_5) \, iS_F(p-w) \, (-2g \gamma_5) \,
 iS_F(k-w) \,  (-2g \gamma_5) \, \frac{i}{w^2}
\end{eqnarray}
which can be simplified to:
\begin{eqnarray}
    \Gamma_A &=&  16 \pi i \alpha^2 \left[ \not \! k \not \! p
    J^{(0)} - (\not \! k \gamma^{\nu} + \gamma^{\nu} \not \! p) J_{\nu}^{(1)}
    + K^{(0)} \right]
\end{eqnarray}
where
\begin{eqnarray}
{\it J}^{(0)}&=&\int \,d^4w\,\frac{1}
{w^2\,(p-w)^2\,(k-w)^2}\\
{\it J}^{(1)}_{\mu}&=&\int \,d^4w\,\frac{w_{\mu}}
{w^2\,(p-w)^2\,(k-w)^2}\\
{\it K}^{(0)}&=&\int \,d^4w\,\frac{1}
{(p-w)^2\,(k-w)^2}
\end{eqnarray}
The exact analytical expressions for these three integrals are known
\cite {BC,KRP}. They involve basic functions of momenta $k$ and
$p$ and a spence function. We believe that it is highly non-trivial 
to construct a non-perturbative vertex which reduces to this complicated 
form in the weak coupling regime. However, asymptotic behaviour
can be reproduced to some extent. Simple power counting reveals that
the integrals $J^{(0)}$ and ${\it J}^{(1)}_{\mu}$ are perfectly
well-behaved in the ultraviolet regime. However, ${\it K}^{(0)}$ is
logarithmically divergent. We now show that our vertex {\em ansatz}
also exhibits this behaviour. Using the fact that perturbatively
${\cal M}(p^2)=0$, we can re-write Eq. (4)  as follows:
\begin{eqnarray}
 F(p^2) &=& 1 \;- \frac{\alpha}{\pi^3} \frac{1}{p^2} \int
                          d^4k \,
                          \frac{F(k^2) F(p^2)}{k^2}\
                          \frac{k \cdot p}{q^2}  
\end{eqnarray}
where we have employed Eq. (6), and the Feynman rule for the vertex. 
Carrying out angular and radial integrations respectively and 
retaining the leading log terms, we get
\begin{eqnarray}
F(p^2) &=& 1 + \frac{\alpha}{2 \pi} \; {\rm ln} \, \frac{p^2}{\Lambda^2}
\end{eqnarray}
Therefore, the proposed vertex ansatz can be written perturbatively as 
follows
\begin{eqnarray}
\Gamma_A(k,p)= \frac{1}{F(k^2)F(p^2)} = 1 + \frac{\alpha}{\pi} \,
  {\rm ln} \, \frac{k^2p^2}{\Lambda^2} + {\cal O}(\alpha^2)
\end{eqnarray}
which is logarithmically divergent in the ultraviolet regime just as
the real vertex to ${\cal O}(\alpha)$. Therefore, perturbatively
our vertex {\em ansatz} is more realistic than the bare vertex.

 An added advantage of 
using the proposed vertex {\it ansatz} is that Eq.(\ref{Meqn})
can be solved independently of Eq.(\ref{Feqn}). Therefore, it serves
a purpose similar to that of Mandlestam's choice \cite{mandle}
of the 3-gluon
vertex in studying the Schwinger-Dyson equation of the gluon
propagator.

As the unknown
functions $F$ and ${\cal M}$ do not depend upon the angle between $k$ and
$p$, we can perform angular integration to arrive at
\vspace {2mm}
\begin{eqnarray}
 F(p^2) &=& 1 \;- \frac{\alpha}{2 \pi} \int
                          dk^2 \,
                          \frac{1}{k^2+{\cal M}^2(k^2)}\
           \left[ \frac{k^4}{p^4} \theta(p^2-k^2)+ \theta(k^2-p^2) \right]
       \label{Feqnr}  \\    
 {\cal M}(p^2)  &=& \frac{\alpha}{\pi} \int 
                          dk^2 \,
                          \frac{{\cal M}(k^2)}{k^2+{\cal M}^2(k^2)}\ 
   \left[ \frac{k^2}{p^2} \theta(p^2-k^2)+ \theta(k^2-p^2) \right]   
               \quad.      \label{Meqnr}
\end{eqnarray}
Such equations are known to have a non-trivial solution for the
mass function above a critical value of the coupling $\alpha=\alpha_c$.
 In the neighbourhood of the
critical coupling, when the generated mass is still small, we can put
${\cal M}^2=0$. Then Eqs. (\ref{Feqnr},\ref{Meqnr}) decouple from
each other completely. The leading log solution for  $F(p^2)$ is
then:
\vspace {2mm}
\begin{eqnarray}
F(p^2) &=& 1 + \frac{\alpha}{2 \pi} \; {\rm ln} \, \frac{p^2}{\Lambda^2}
\qquad.
\end{eqnarray}
As for the mass function, multiplicative renormalizability demands a 
solution of the type $M(p^2)\simeq (p^2)^{-s}$. Substituting this in
Eq. (\ref{Meqnr}) and performing radial integration, we find 
\vspace {2mm}
\begin{eqnarray}
  s &=&  \frac{1}{2} \; \pm   \frac{1}{2} \; \sqrt{1 - 
\frac{\alpha}{\alpha_c} }
\end{eqnarray}
where $\alpha_c=\pi/4$. For $\alpha > \alpha_c$ the solution of the
mass function enters the complex plane indicating that a phase transition
has taken place from perturbative to non-perturbative solution 
corresponding to the dynamical generation of mass. Numerically, above
$\alpha_c$, 
we solve Eq. (\ref{Meqnr}) in a two-step process. We first use the iterative 
method to get close to the solution and then refine the answer by
converting the integral equation into a set of simultaneous nonlinear
equations to be solved by Newton-Raphson method. 
In Fig. 3., we
have drawn the Euclidean mass $M$(which can be taken to be ${\cal M}(0)$)
as a function of the coupling $\alpha$. 
We see that it obeys Miransky scaling law and can be fitted
 to the form
\vspace {2mm}
\begin{eqnarray}
  \frac{M}{\Lambda} &=& {\rm exp} \, \left[ - \frac{A}{ \sqrt { 
\frac{\alpha}{\alpha_c} - 1  } } + B \right]
\end{eqnarray}
very well by the choice $A=0.97 \pi$ and $B=1.45$. These numbers are
close to the ones found in \cite{mike1} although the value of the
critical coupling is of course different. The slight mismatch
in the values of $A$ and $B$ is
due to the fact that in the logarithmic grid of momenta, we choose
30 points per decade and do not extrapolate the result to an infinite
number of points, an exercise carried out in \cite{mike1}.
We also compute $F(p^2)$ for various values of $\alpha$  and find
that the closer we approach $\alpha_c$, where the generated mass is still 
small, starting from a larger value of $\alpha$,
the numerical result gets closer and closer to the analytical result as 
expected, Fig. 4.


\section{The Wess-Zumino Model}

We now extend the particle spectrum by doubling the number of scalars
to discuss the massless Wess-Zumino model, characterized by the
following Lagrangian:
\vspace {2mm}
\begin{eqnarray}
{\cal{L}} = \frac{1}{2} (\partial_{\mu} A)^2 + 
            \frac{1}{2} (\partial_{\mu} B)^2 
          +  \frac{1}{2} (i\bar{\psi} \gamma^{\mu} \partial_{\mu} \psi)
 -\frac{1}{2} g^2 (A^2 + B^2)^2 - g\bar{\psi} (B - i A \gamma_5) \psi \;.
\end{eqnarray}
The Schwinger-Dyson equation for the fermion propagator in this model is 
depicted in Fig. 5. Before we embark on solving this equation, 
we must re-address the validity of the ansatz for the 3-point vertex
which we made for the Yukawa case. To the one loop level in 
perturbation theory, one now has contributions from both the scalars
as depicted in Fig. 6. Therefore, e.g., for scalar $A$, we can write
\begin{eqnarray}
 -2g \gamma_5 \Gamma_A &=& -2g \gamma_5 + \int \frac{d^4 w}{(2 \pi)^4}
 \, (-2g \gamma_5) \, iS_F(p-w) \, (-2g \gamma_5) \,
 iS_F(k-w) \,  (-2g \gamma_5) \, \frac{i}{w^2}  \nonumber  \\
 &+& \int \frac{d^4 w}{(2 \pi)^4}
 \, (-2ig ) \, iS_F(p-w) \, (-2g \gamma_5) \,
 iS_F(k-w) \,  (-2ig) \, \frac{i}{w^2} = 0 \quad.
\end{eqnarray}
The same is the case for $\Gamma_B$, i.e., 
in the presence of both the scalars $A$ and $B$, none
of the 3-point vertices gets modified at ${\cal O}(\alpha)$ in perturbation
theory. Therefore, in the Wess-Zumino case, it is more reasonable to use
the bare vertex instead of the {\it ansatz} earlier made. Though
one would now expect to solve coupled integral equations for 
$F(p^2)$ and ${\cal M}(p^2)$, a miraculous cancellation of terms
takes place as evident from the following Schwinger-Dyson equation
for the fermion propagator:
\vspace {2mm}
\begin{eqnarray}
\nonumber 
\frac{\not \! p- {\cal M}(p^2)}{iF(p^2)} &=&
\frac{\not \! p}{i} - \frac{\alpha}{\pi^3}   \; \int 
d^4 k  \left[ \frac{F(k^2)}{\not \! k- {\cal M}(p^2) } \; \frac{1}{q^2} 
\right] +  \frac{\alpha}{\pi^3}   \; \int 
d^4 k  \left[ \gamma_5  \frac{F(k^2)}{\not \! k- {\cal M}(p^2) } 
 \gamma_5 \; \frac{1}{q^2} 
\right] \; .  \\
\end{eqnarray}
Taking the trace of this equation, we get 
\vspace {2mm}
\begin{eqnarray*}
 {\cal M}(p^2)&=&0 \qquad.
\end{eqnarray*}
As the cancellation of terms takes place at the very beginning,
it is easy to see that dynamical mass generation will remain an 
impossibility for the full vertex and the full scalar propagator
as long as they are identical for both the scalars. The vertex 
corrections for $A$ and $B$ have been proven to be equal
up to ${\cal O}(\alpha^2)$ \cite{jjw}. We shall shortly see
that the same is true for the full scalar propagator at least up to 
${\cal O}(\alpha)$. This is in accordance with 
the arguments based on non-renormalization theorem.  
SUSY plays a role in providing same number of 
bosonic and fermionic degrees of freedom. We have seen from the
case of the Yukawa Model that without this equality, it will not be 
possible to prevent dynamical
mass generation. Secondly, SUSY imposes relations on couplings
of the two scalars with the fermions. This relationship is crucial
in preventing dynamical mass generation.

\noindent
As far as wavefunction renormalization $F(p^2)$  is concerned, its
leading log behaviour gets modified slightly, by the inclusion of the
other scalar, to:
\vspace {2mm}
\begin{eqnarray}
F(p^2) &=& 1 + \frac{\alpha}{ \pi} \; {\rm ln} \, \frac{p^2}{\Lambda^2}
\quad.
\end{eqnarray}
Although it is an interesting conclusion in its own right  that
supersymmetry prevents dynamical mass generation for fermions in the
Wess-Zumino model, another important issue to
probe will be whether supersymmetry itself remains intact, i.e., whether
the scalars can also be kept massless. This is what we discuss now.
The Schwinger-Dyson equation for the scalar (for example $A$) has been
depicted in Fig. 7. A scalar propagator, unlike a fermion, needs only
one unknown function to describe it. But we shall prefer to split it into
two parts and write the full scalar propagator as follows:
\vspace {2mm}
\begin{eqnarray}
    S_{A}(p)&=& \frac{F_A(p^2)}{p^2 - {\cal M}_A^2(p^2)}\qquad .
\end{eqnarray}   
The non-zero value of the mass function ${\cal M}_A(p^2)$ will be
responsible for shifting the pole from $p^2=0$ to some finite value,
generating the mass for the scalar dynamically. $F_A(p^2)$ on the other
hand is the scalar wavefunction renormalization. The SD-equation for 
the scalar propagator in Euclidean space can now be written as:
\vspace {2mm}
\begin{eqnarray}
\nonumber
  \frac{p^2+{\cal M}_A^2(p^2)}{F_A(p^2)}  &=& p^2 + 
          \frac{3\alpha}{2\pi^3} \int 
          d^4k \, \frac{F_A(k^2)}{k^2+{\cal M}_A^2(k^2)} + 
	 \frac{\alpha}{2\pi^3} \int 
          d^4k \, \frac{F_B(k^2)}{k^2+{\cal M}_B^2(k^2)}  \\
	&& -   \frac{2\alpha}{\pi^3} \int 
          d^4k \, \frac{F(k^2)F(q^2)}{k^2 q^2} \; \Gamma_A(k,p) \;
	    k \cdot q 
\end{eqnarray}
where we have used the fact that the fermions do not acquire mass. 
If we want to preserve supersymmetry and do not want the scalars to
acquire mass, we must have:
\vspace {2mm}
\begin{eqnarray}
   {\cal M}_A(p^2)={\cal M}_B(p^2)=0   \qquad.
\end{eqnarray}
We are then left with:
\vspace {2mm}
\begin{eqnarray}
  \frac{1}{F_A(p^2)}  &=& 1 + 
          \frac{\alpha}{2\pi^3p^2} \int 
          \, \frac{d^4k}{k^2} 
	   \left[ 3F_A(k^2)  + F_B(k^2) - 
    4 \frac{ k \cdot q }{q^2} \; F(k^2)F(q^2) \Gamma_A(k,p) \right]
\end{eqnarray}
and there is a similar equation for the scalar $B$:
\vspace {2mm}
\begin{eqnarray}
  \frac{1}{F_B(p^2)}  &=& 1 + 
          \frac{\alpha}{2\pi^3p^2} \int 
          \, \frac{d^4k}{k^2} 
	   \left[ 3F_B(k^2)  + F_A(k^2) - 
    4 \frac{ k \cdot q }{q^2} \; F(k^2)F(q^2) \Gamma_B(k,p) \right]
     \;.
\end{eqnarray}
These equations should yield a solution for $F_A(p^2)$ and  $F_B(p^2)$
such that it does not change the position of the pole 
for the scalar propagator and that the quadratic divergences cancel.
It is well-known that it does happen in perturbation theory to 
${\cal O}(\alpha)$. In fact, one can evaluate $F_A(p^2)$ and 
$F_B(p^2)$. The leading log expression for these functions to 
${\cal O}(\alpha)$ is
\vspace {2mm}
\begin{eqnarray}
F_A(p^2)= F_B(p^2) = 1 + \frac{\alpha}{ \pi} \; {\rm ln} \, 
\frac{p^2}{\Lambda^2}
\end{eqnarray}
which is exactly the same expression as that for $F(p^2)$ for the
fermion propagator. This result indicates that supersymmetry need 
not be broken.

\section{Conclusions}

We have studied the Schwinger-Dyson equations for the Yukawa 
(a scalar interacting with a fermion with a $\gamma_5$ type
interaction) and the Wess-Zumino models. In the simple Yukawa 
model, we propose a vertex $\it ansatz$ which we argue should 
perform better than the bare vertex. In the 
quenched approximation, 
we find dynamical mass generation for fermions above a critical
value of the coupling $\alpha_c=\pi/4$. The generated Euclidean mass obeys
Miransky scaling. When we extend this Yukawa model to equate the
scalar and fermionic degrees of freedom (Wess-Zumino model), 
we find that a neat cancellation of terms occurs and there is
no mass generation for the fermions. This fact remains true beyond the
rainbow approximation and is supported by perturbative calculations
available for the 3-point vertex to ${\cal O}(\alpha^2)$ and of the 
scalar propagator.
This result was expected on the basis of non-renormalization theorem.
The two approaches will remain in agreement provided the full vertex and 
the full scalar propagator
as long as they are identical for both the scalars to higher orders in
perturbation theory as well.

If supersymmetry has to be preserved, the scalars should also acquire no
mass dynamically. Studying the Schwinger-Dyson equations for the
scalars, we observe that such a solution is allowed and in fact
leads to the wavefunction renormalization function for the
scalars which is exactly the same as that for the fermion.

It is more interesting to see the role of supersymmetry in 
more complicated theories such as SQED. The studies so far carried out 
in superfield and component formalism seem to arrive at different
conclusions. We plan to present our work in this context in a future 
publication.

   \vskip 1cm
   
\noindent{\bf Acknowledgements}
 
   \vskip 5mm
This work was partly supported by a TWAS-AIC award and CONACYT-SNI 
(M\'exico).
AB is also grateful for the hospitality of Instituto de F{\'\i}sica, 
Benem\'erita Universidad Aut\'onoma de Puebla  (BUAP) during his stay 
there, where a part of the work was done.

\hsize=16.5cm
\baselineskip=6mm

\vfil\eject
\vskip 1cm
\noindent{\bf Figure Captions}
\vskip 5mm
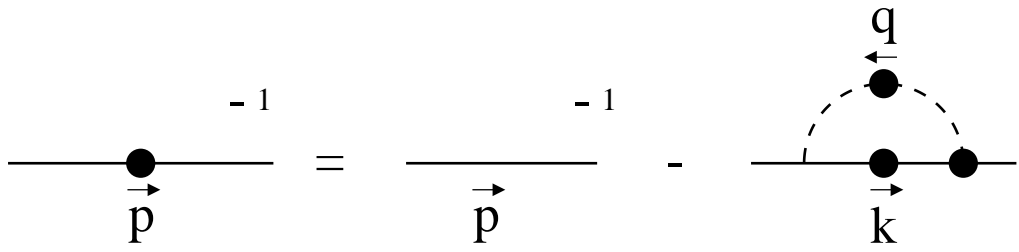
\noindent{\bf Fig. 1.} Schwinger-Dyson equation for the fermion propagator
in the Yukawa model.
The solid lines represent fermions and the dashed the scalar.  The solid 
dots indicate full, as opposed to bare, quantities.
\vskip 5mm
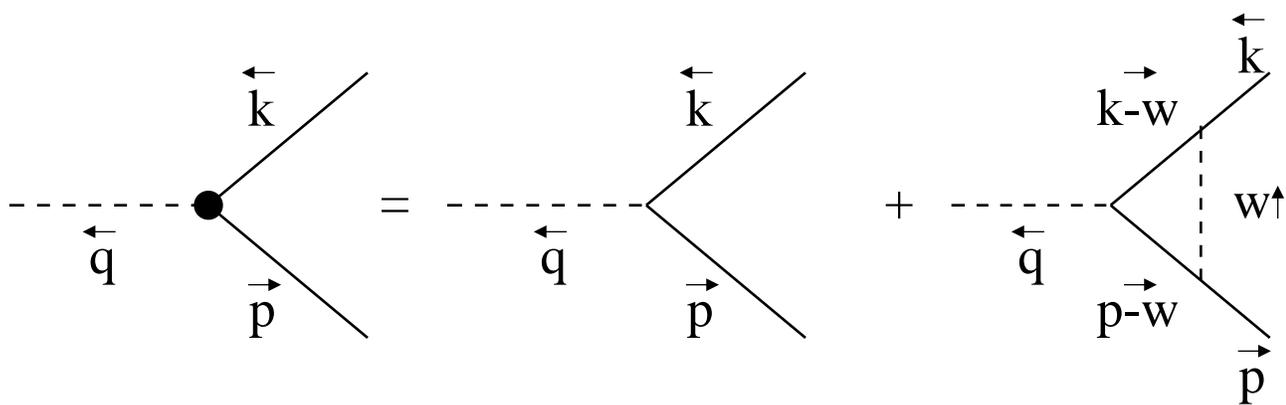
\noindent{\bf Fig. 2.}
One loop perturbative expansion for the vertex in the Yukawa model.
\vskip 5mm
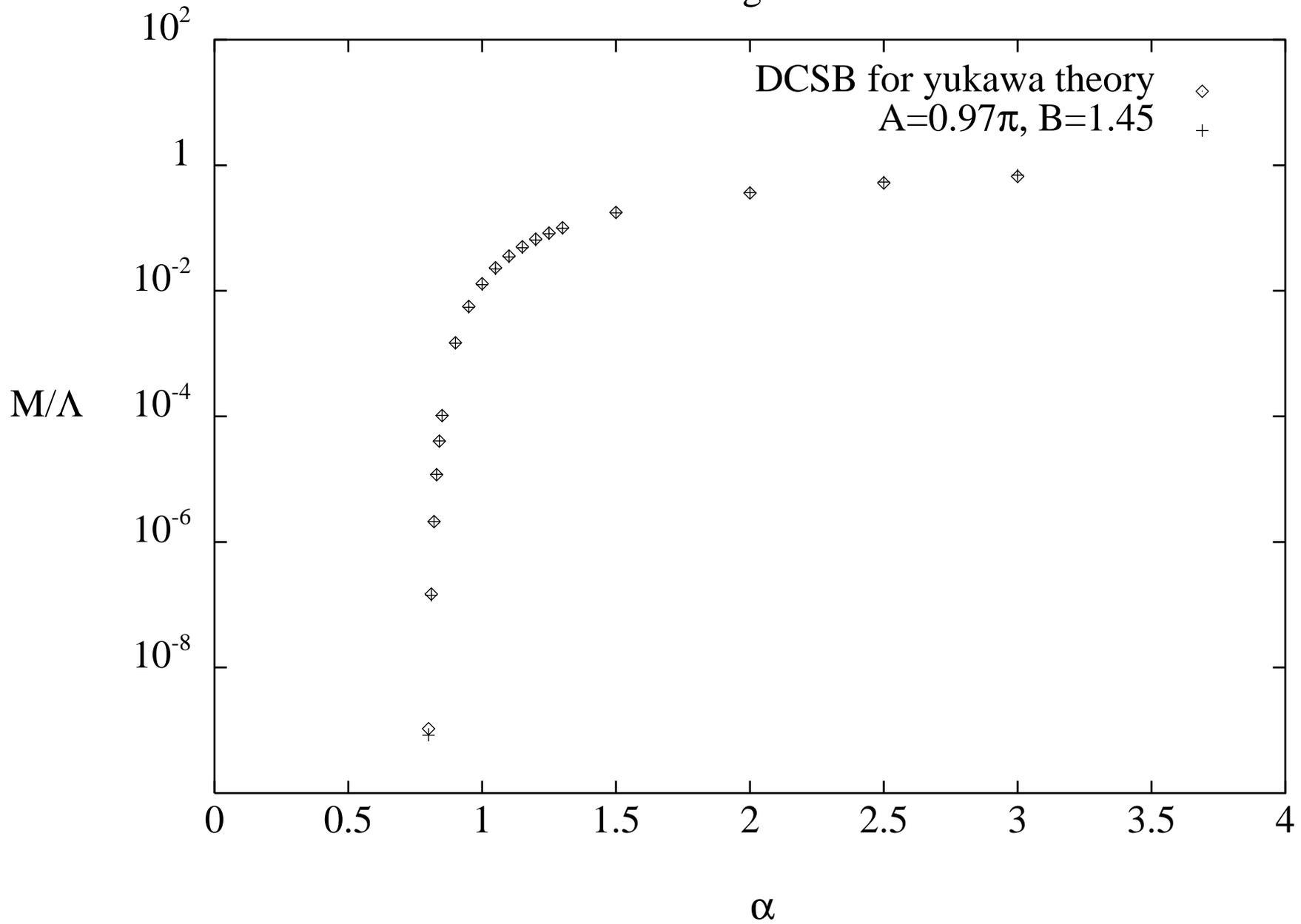
\noindent{\bf Fig. 3.}
The dynamically generated mass $(M/\Lambda)$ versus the 3-point coupling
$\alpha$ in the Yukawa model. The critical coupling is $\alpha_c=\pi/4$, 
above which the mass can be seen to be bifurcating away from the
chirally symmetric solution. $\diamond$s represent the numerical result
and $+$s the numerical fit to the form 
$ M = \Lambda {\rm exp} \, \left[ - A / \sqrt { 
{\alpha}/{\alpha_c} - 1  }  + B \right] $ with $A=0.97 \pi$ and $B=1.45$.
\vskip 5mm
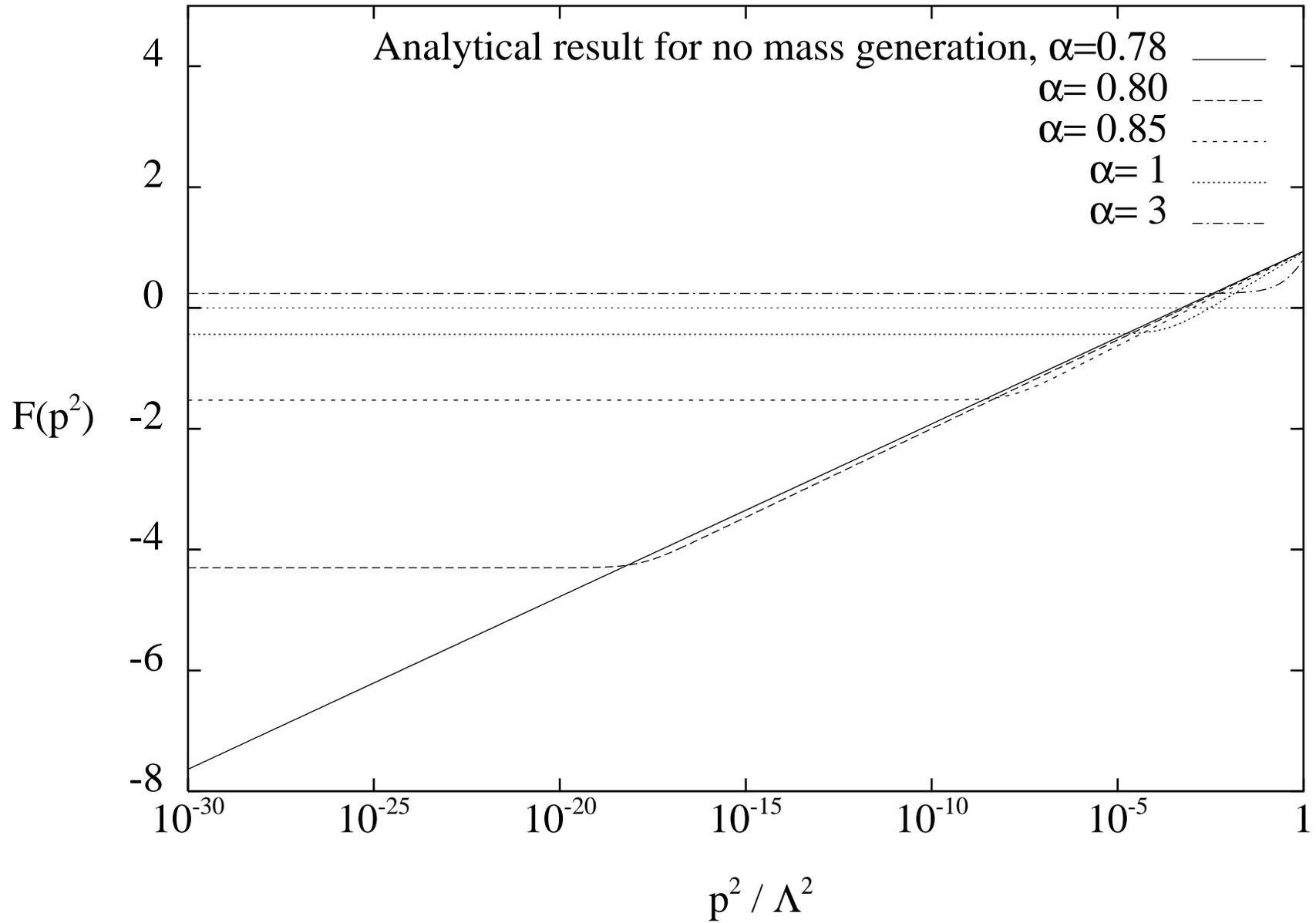
\noindent{\bf Fig. 4.}
The wavefunction renormalization function $F(p^2)$ in the Yukawa model
for various values of the coupling $\alpha$. The solid line corresponds 
to the analytical expression $ 1 - \alpha/{4 \pi} + (\alpha/{2 \pi})
\; {\rm ln} (p^2/\Lambda^2)$ in case of no mass generation for
$\alpha=0.78$.
\vskip 5mm
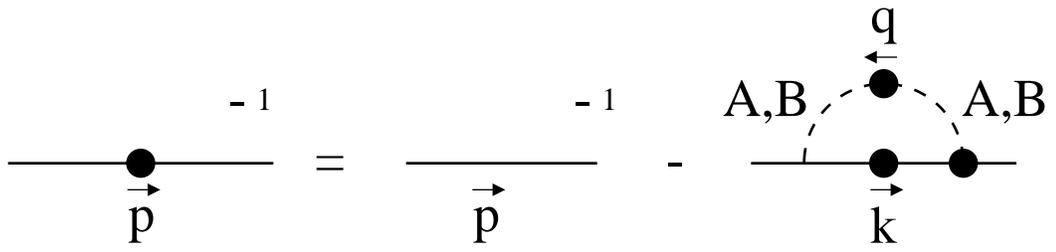
\noindent{\bf Fig. 5.}
Schwinger-Dyson equation for the fermion propagator
in the Wess-Zumino model.
\vskip 5mm
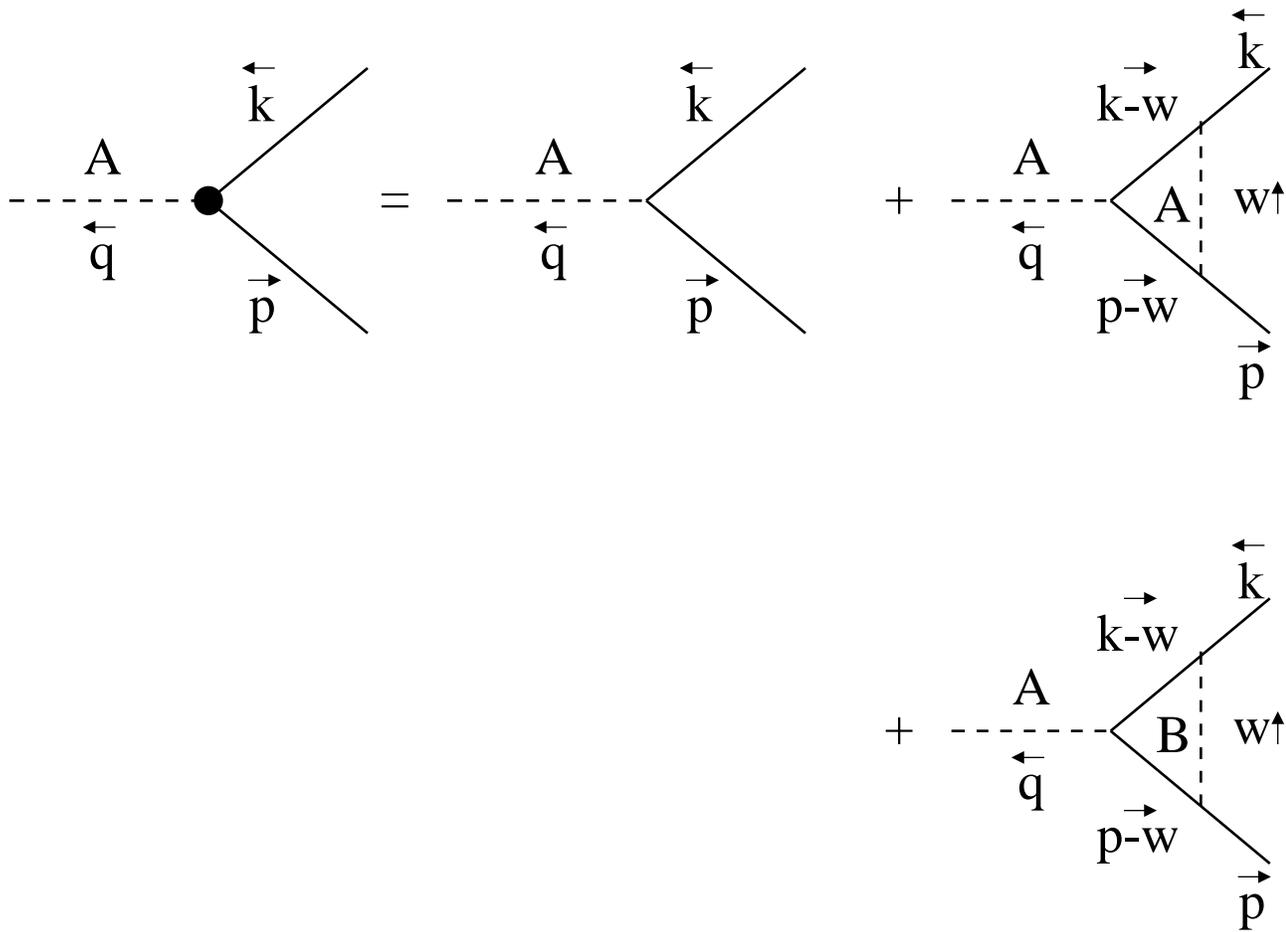
\noindent{\bf Fig. 6.}
One loop perturbative expansion for the vertex in the Wess-Zumino
model.
We have shown the case for scalar $A$. A similar diagram exists for 
scalar $B$.
\vskip 5mm
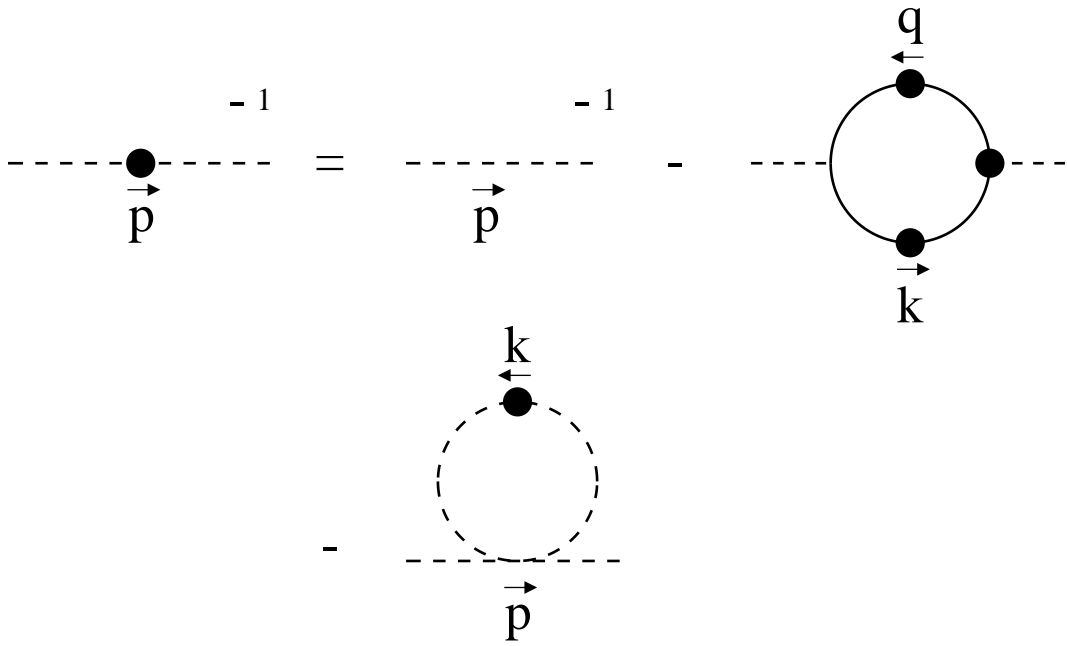
\noindent{\bf Fig. 7.}
Schwinger-Dyson equation for the scalar propagator
in the Wess-Zumino model.


\begin{thebibliography}{999}
\bibitem{wess} J. Wess and B. Zumino, Nucl. Phys. {\bf B70} 39 (1974);
Phys. Lett. {\bf B49} 52 (1974)
\bibitem{roberts} For a review see: C.D. Roberts and A.G. Williams, Prog. 
Part. Nucl. Phys.
{\bf 33} 477 (1994). 
\bibitem{miransky} V.A. Miransky, Nuovo Cim. {\bf 90A} 149 (1985)~;
 Sov. Phys. JETP {\bf 61} 905 (1985)~;\\
P.I. Fomin, V.P. Gusynin, V.A. Miransky and Yu.A. Sitenko, La rivista
del Nuovo Cim. {\bf 6}, numero5, 1 (1983).
\bibitem{mike} D.C. Curtis and M.R. Pennington, Phys. Rev. {\bf D42}
4165 (1990).
\bibitem{dong} Z. Dong, H.J. Munczek and C.D. Roberts, Phys. Lett. {\bf B333}
536 (1994).
\bibitem{bashir} A. Bashir and M.R. Pennington, Phys. Rev. D50 7679 (1994).
\bibitem{clark} T.E. Clark and S.T. Love, Nucl. Phys. {\bf B310} 371 (1988)
\bibitem{kaiser} A. Kaiser and S.B. Selipsky, hep-th/9708087, Yale preprint
YCTP-P14-97.
\bibitem{apple} T. Appelquist, A. Nyffeler and S.B. Belipsky, hep-th/9709177,
Yale preprint YCTP-P12-97.
\bibitem{BC} J.S. Ball and T.-W. Chiu, Phys. Rev. {\bf D22}, 2542 (1980).
\bibitem{KRP} A. K{\i}z{\i}lers\"u, M. Reenders and M.R. Pennington, Phys.
Rev. {\bf D52} 1242 (1995).
\bibitem{mandle} S. Mandlestam, Phys. Rev. {\bf D20} 3223 (1979).
\bibitem{mike1} D.C. Curtis and M.R. Pennington, Phys. Rev. {\bf D48}
4933 (1993).
\bibitem{jjw} I. Jack, D.R.T. Jones and P. West, Phys. Lett. {\bf B258}
382 (1991).
\end{thebibliography}
\end{document}